\documentclass{aa}  %[referee]
\usepackage{natbib}
\bibpunct{(}{)}{;}{a}{}{,} % to follow the A&A style
\usepackage{graphicx}
\usepackage{txfonts}
\usepackage[colorlinks=true, allcolors=blue]{hyperref}
\newcommand{\cutting}{\alpha_{C}}
\newcommand{\hypot}{\alpha_{H}}
\newcommand{\alignT}{\phi_{A}}
\newcommand{\alignN}{\theta_{A}}
\newcommand{\alignsymbol}{\hat{\vec{A}}}
\newcommand{\examplestream}{\rm{E2S4}}

\begin{document}

   \title{Statistical analysis of orientation, shape, and size of solar wind switchbacks}

   %\subtitle{I. Overviewing the $\kappa$-mechanism}

   \author{R. Laker\inst{1}\thanks{Corresponding author: Ronan Laker \email{ronan.laker15$@$imperial.ac.uk}}
   \and
   T. S. Horbury\inst{1}
   \and
   S. D. Bale\inst{1,2,3}
   \and
   L. Matteini\inst{1}
   \and
   T. Woolley\inst{1}
   \and
   L. D. Woodham\inst{1}
   \and
   S. T. Badman\inst{2,3}
   \and
   M. Pulupa\inst{3}
   \and
   J. C. Kasper\inst{4,5}
   \and
   M. Stevens\inst{4}
   \and
   A. W. Case\inst{4}
   \and
   K. E. Korreck\inst{4}
   }

   \institute{Imperial College London, South Kensington Campus, London, SW7 2AZ, UK
   \and
   Physics Department, University of California, Berkeley, CA 94720-7300, USA
   \and
   Space Sciences Laboratory, University of California, Berkeley, CA 94720-7450, USA
   \and
   Smithsonian Astrophysical Observatory, Cambridge, MA 02138, USA
   \and
   Climate and Space Sciences and Engineering, University of Michigan, Ann Arbor, MI 48109, USA
   }

   \date{Received XXXX; accepted YYYY}

% \abstract{}{}{}{}{} 
% 5 {} token are mandatory
 
  \abstract
  % context heading (optional)
  % {} leave it empty if necessary  
   {One of the main discoveries from the first two orbits of Parker Solar Probe (PSP) was the presence of magnetic switchbacks, whose deflections dominated the magnetic field measurements. Determining their shape and size could provide evidence of their origin, which is still unclear. Previous work with a single solar wind stream has indicated that these are long, thin structures although the direction of their major axis could not be determined}
  % aims heading (mandatory)
   {We investigate if this long, thin nature extends to other solar wind streams, while determining the direction along which the switchbacks within a stream were aligned. We try to understand how the size and orientation of the switchbacks, along with the flow velocity and spacecraft trajectory, combine to produce the observed structure durations for past and future orbits.}
  % methods heading (mandatory)
   {The direction at which the spacecraft cuts through each switchback depended on the relative velocity of the plasma to the spacecraft and the alignment direction for that stream. We searched for the alignment direction that produced a combination of a spacecraft cutting direction and switchback duration that was most consistent with long, thin structures. The expected form of a long, thin structure was fitted to the results of the best alignment direction, which determined the width and aspect ratio of the switchbacks for that stream.
   }
  % results heading (mandatory)
   {We find that switchbacks consistently demonstrate a non-radial alignment in the same sense as the Parker spiral field, but not the background flow direction within each stream. This alignment direction varied between streams. The switchbacks had a mean width of $50,000 \, \rm{km}$, with an aspect ratio of the order of $10$.}
   %add in stuff about what this means for generation mechanisms here -but have to decide what that is first
  % conclusions heading (optional), leave it empty if necessary 
   {We conclude that switchbacks are not aligned along the background flow direction, but instead aligned along the local Parker spiral, perhaps suggesting that they propagate along the magnetic field. Since the observed switchback duration depends on how the spacecraft cuts through the structure, the duration alone cannot be used to determine the size or influence of an individual event. For future PSP orbits, a larger spacecraft transverse component combined with more radially aligned switchbacks will lead to long duration switchbacks becoming less common.}


   \maketitle
%
%-------------------------------------------------------------------

\section{Introduction}
%go over introduction and add in stuff you said in ESA switchback section
Parker Solar Probe (PSP) began a new era for heliophysics when it surpassed the Helios probes as the closest spacecraft to measure the solar wind at $35$ solar radii from the Sun. \citet{Bale2019} and \citet{Kasper2019} report the dominance of ‘switchbacks’ in the magnetic field, discrete large angular deflections from the background magnetic field direction lasting from seconds to hours \citep{DudokdeWit2020}. Being Alfv\'enic in nature \citep{Matteini2014a}, these switchbacks are associated with an increase in plasma velocity and therefore are an important contribution to the overall mass flux of the solar wind.

%The increased fluctuation amplitude (deltaB/B = 1) is caused by an Alfv\'en speed and field magnitude that comes with PSP superior proximity to the Sun,
Switchbacks have previously been observed with data from Helios \citep{Horbury2018a} and Ulysses \citep{Balogh1999, Neugebauer2013}, but it was the magnitude and prevalence of the switchbacks that made the PSP observations striking. Switchbacks have been shown to be consistent with folds in the magnetic field, rather than reversals in local polarity \citep{Balogh1999, Mcmanus2020}. \citet{Horbury2020a} describe the magnetic field vectors within switchbacks as being arc-polarised rotations on a sphere of constant $|\vec{B}|$, while also showing that the direction of switchback deflection was broadly consistent on the timescales of hours to days. 

The origin of these structures is still unclear, whether they are a product of fluctuations expanding with the solar wind \citep{Squire2020a}, or the remnants of transient events originating in the solar corona \citep{Horbury2020a, Sterling2020}. One such example of a transient coronal event are coronal jets, which have been studied in extreme ultraviolet (EUV) and X-ray observations \citep{Raouafi2016}. \citet{DudokdeWit2020} demonstrate, with waiting time statistics, that the switchbacks are correlated and exhibit a long term memory, which supports the idea of a coronal origin.
\begin{figure*}[ht!]
\centering
\includegraphics[width=\textwidth]{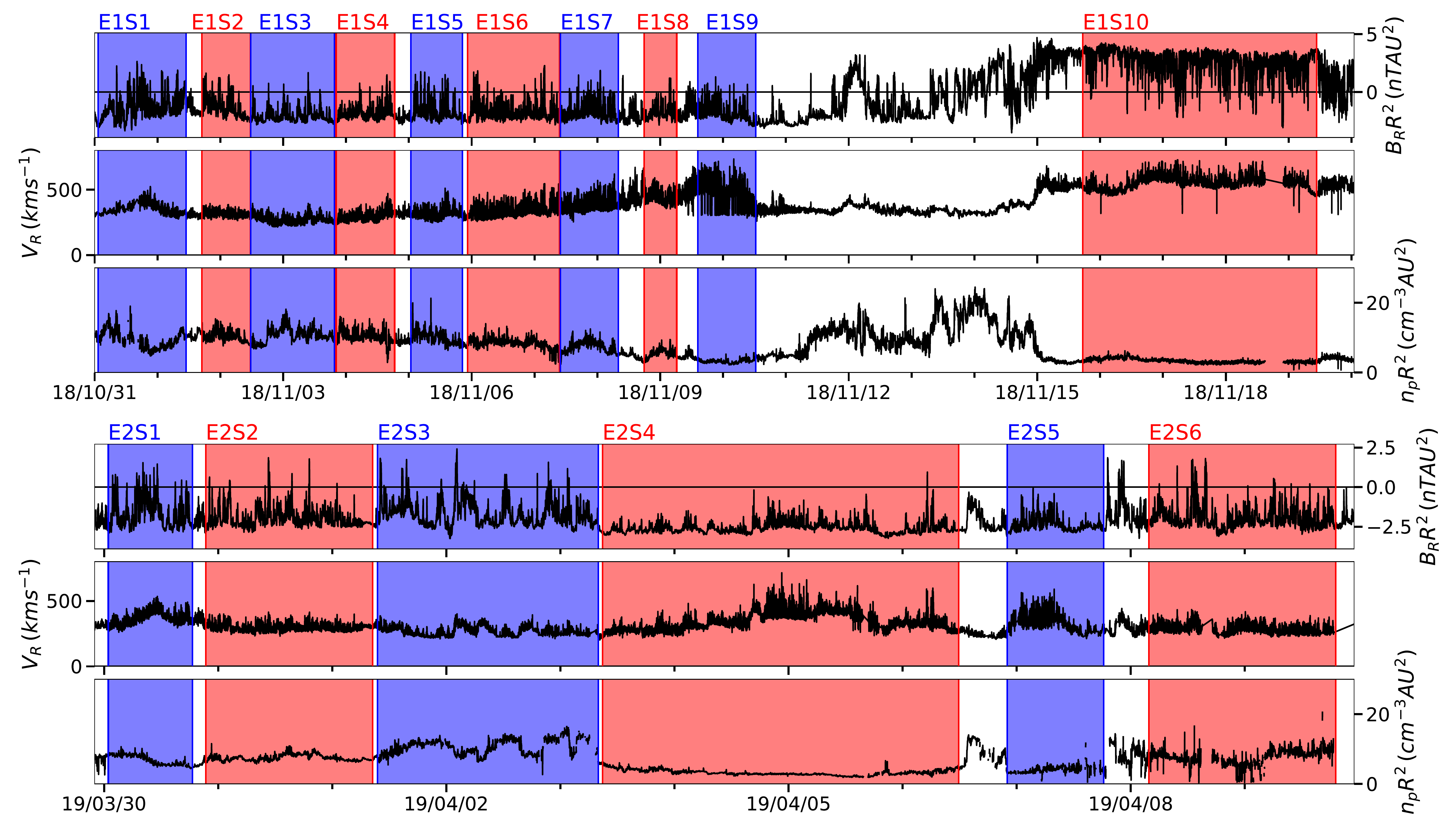}
\caption{Summary of solar wind streams for encounter 1 and 2, along with their identification numbers. All streams in encounter 1 (except $E1S10$) are from the same equatorial coronal hole. The period between $E1S9$ and $E1S10$ was ignored, as this is a heliospheric current sheet crossing \citep{Szabo2020}. Encounter 2 streams were picked according to the \citet{Rouillard2020a} classification of non and streamer belt solar wind. A more detailed summary of these streams can be found in Table \ref{tab:1}. The top panels for both encounters show $B_{R}$ multiplied by the radial distance to the Sun squared. This is an attempt to remove the first order variation in $B_{R}$, normalising the switchback amplitude with distance.}\label{fig:streamsummary}
\end{figure*}
A key assumption of the coronal origin theory is that fluctuations can survive out to the distances of PSP. This has been shown to be possible with recent magnetohydrodynamic (MHD) simulations when $|B|$ is constant \citep{Tenerani2020}, but it is unclear whether this extends to larger distances, such as the Ulysses observations \citep{Neugebauer2013}. Although switchbacks are the dominant signal in PSP’s first encounter, they are not ubiquitous, occurring in ‘patches’ separated by a ‘quiet’ radial field \citep{Bale2019, Horbury2020a}. These quiet regions exhibit a wide array of kinetic wave activity that is not seen in the patches \citep{Verniero2020, Bowen2020}. This combined with differences in magnetic compressibility could indicate that patches of switchbacks are impulsive events overlain on a fundamentally different background plasma \citep{Woodham2020}. This observation may prove to be important in determining the origin of switchbacks, as it is not yet clear whether each switchback is an isolated object, or if a patch of switchbacks are the sampling of a larger physical structure.

\citet{Horbury2020a} provided the first piece of evidence of the 3D spatial structure of these switchbacks, using a perihelion stream to determine their structure as long and thin, with a high aspect ratio in roughly the radial direction. In this paper, we extend this idea to multiple solar wind streams. For the first encounter, PSP was connected to a small over-expanded coronal hole from the $29^{th}$ October 2018 to the $13^{th}$ November 2018, where the spacecraft then crossed the heliospheric current sheet and connected to a larger coronal hole between $14^{th}$ November 2018 and $23^{rd}$ November 2018 \citep{Badman2020}. We have split the first encounter into ten distinct solar wind streams, based on these patches and other enhancements in density and velocity. These streams are shown in Fig \ref{fig:streamsummary}, and will be referenced by the identification numbers shown above.

Global context for the second encounter with respect to the Sun’s structure was inferred using white light images, which revealed that PSP sampled higher density solar wind when above the streamer belt \citep{Rouillard2020a}. Therefore, the streams for encounter 2, as shown in Fig \ref{fig:streamsummary}, are based on this classification. The presence of switchbacks in both encounters therefore demonstrated that they exist in both slow and fast coronal hole streams as well as within and outside streamer belt flows. 

We demonstrate that an average alignment direction can be found for switchbacks within each solar wind stream. We then compare the alignment direction for each stream to relevant physical directions, including the background flow and local Parker spiral \citep{Parker1958}. Section \ref{sec:method} presents a theoretical discussion of how different physical properties combine to give a range of angles that the spacecraft can cut through the structure. This is followed by a discussion of how the average alignment direction was determined. Finally, the results are presented in Section \ref{sec:results}, and their implications are discussed.

\section{Method}\label{sec:method}

In order to build up a 3D shape of a structure, it would be preferable to obtain multiple images from known angles to the structure in question. This technique has been applied to images of coronal mass ejections \citep{DeKoning2009, Liewer2009}, but cannot be employed in the study of switchbacks due to their largely incompressible nature, meaning they would not be visible in images. Therefore, we are limited to single point spacecraft measurements, such as those provided by PSP during its first two encounters. An average shape can still be investigated from such observations, by studying how the apparent duration in the time series, $t$, varies with the direction at which the spacecraft cuts through the structure. For example, a 2D circle will have the same measured duration when cut through its centre, no matter which direction it is cut from, but an ellipse will have a larger measured duration when cut along its major axis. Inevitably, there will be a range of measured durations from the same cutting angle, as the structure can be sampled off centre, and the structures will not be perfectly uniform in size. Such considerations mean that analysis of their shape must be statistical in nature.
 
\begin{figure}[ht!]
\centering
\includegraphics[width=\hsize]{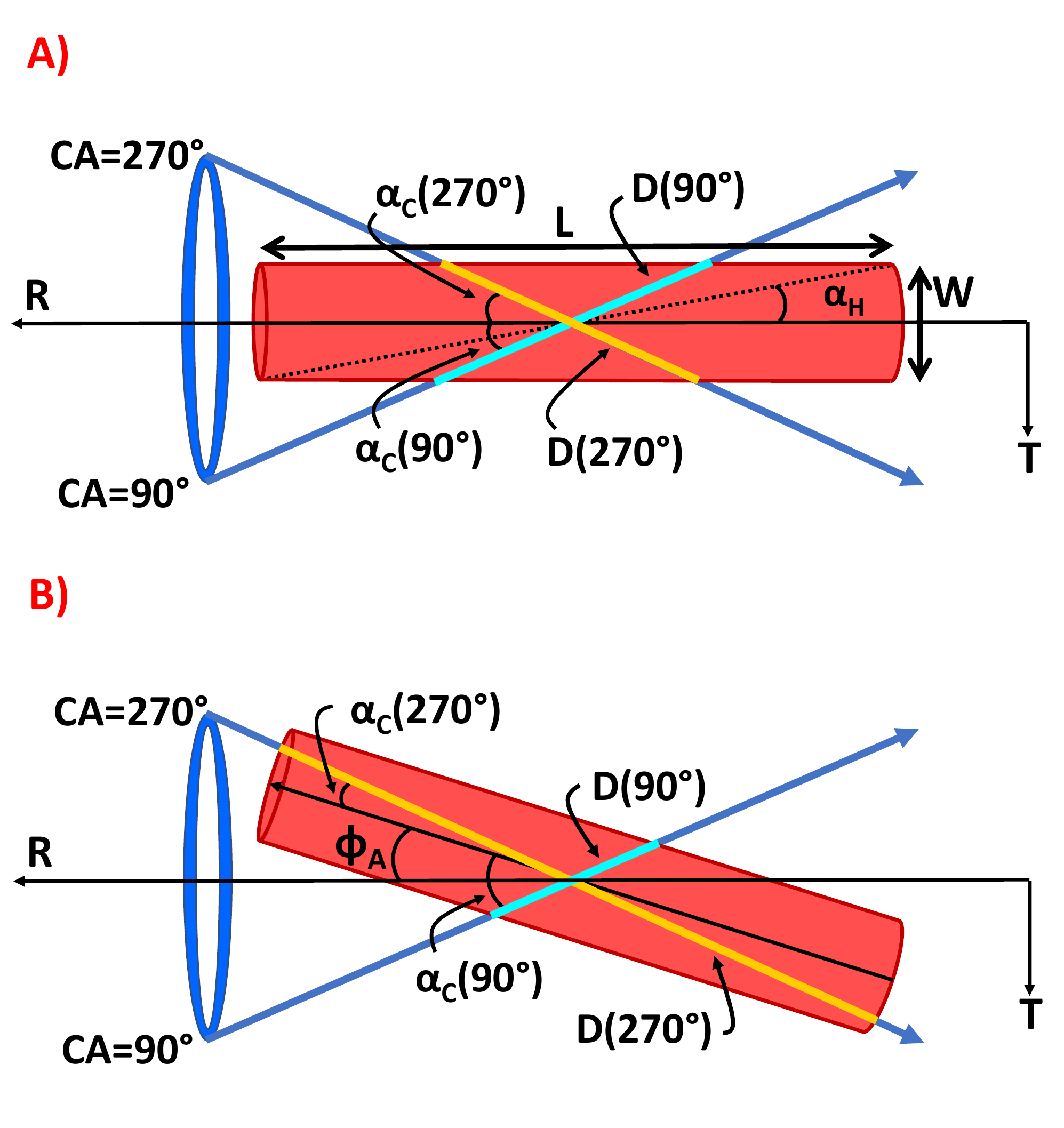}
\caption{A) Example of two spacecraft cuts at a Clock Angle (CA) of $90\degr$ and $270\degr$ when viewed from the N direction. The ellipse represents the ring of possible velocities of the spacecraft from Fig \ref{fig:circle}. In this example, the switchback structure (red cylinder) is centred on the ring of velocities, so both spacecraft cuts are at the same angle, and therefore only one value of $D$ is measured. B) In this example, the alignment direction of the switchbacks has been rotated by $\alignT$ in the RT plane. This means that the spacecraft measures different values of D depending on the CA.}
          \label{fig:cuts}%
\end{figure}

We assume a cylindrically symmetric long, thin structure, as suggested by \citet{Horbury2020a}, aligned along an alignment direction, $\alignsymbol$. The structure has width $W$ and length, $L$, that can be seen in Fig \ref{fig:cuts}. The angle of the hypotenuse, $\hypot$, is then defined as $\arctan\frac{W}{L}$, and gives a measure of the aspect ratio. The distance measured, $D$, by the spacecraft cutting through the centre at an angle, $\cutting$ (cutting angle), is given by:
\begin{equation}\label{e:duration}
D = t \times |\vec{V_{rel}}| = \left\{\begin{matrix}
\frac{W}{\sin\cutting}, & \cutting \geq \hypot\\ \\
\frac{W}{\tan\hypot\cos\cutting}, & \cutting < \hypot.
\end{matrix}\right.
\end{equation}

This analysis requires knowledge of the direction at which the spacecraft passed through each switchback. This can be estimated by first defining the velocity of the plasma relative to the spacecraft:
\begin{equation}
    \vec{V_{rel}} = \left \langle \vec{V_{pl}}  \right \rangle - \vec{V_{sc}},
\end{equation}
where $\left \langle \vec{V_{pl}}  \right \rangle$ is the mean plasma velocity within the switchback relative to the Sun, using proton moments from the Solar Probe Cup \citep{Kasper2016, Case2020}. We use magnetic field measurements from the FIELDS instrument suite \citep{Bale2016}. The spacecraft velocity in Heliocentric Inertial frame, $\vec{V_{sc}}$, has a large +T component at PSP’s perihelion in the Radial-Tangential-Normal (RTN) coordinate system. In these coordinates, R points from the Sun to the spacecraft, N is the component of the solar north direction perpendicular to R, and T completes the right handed set. The cutting angle, $\cutting$, is defined as the angle between $\vec{V_{rel}}$ and $\alignsymbol$. 

\begin{figure}[h!]
\centering
\includegraphics[width=\hsize]{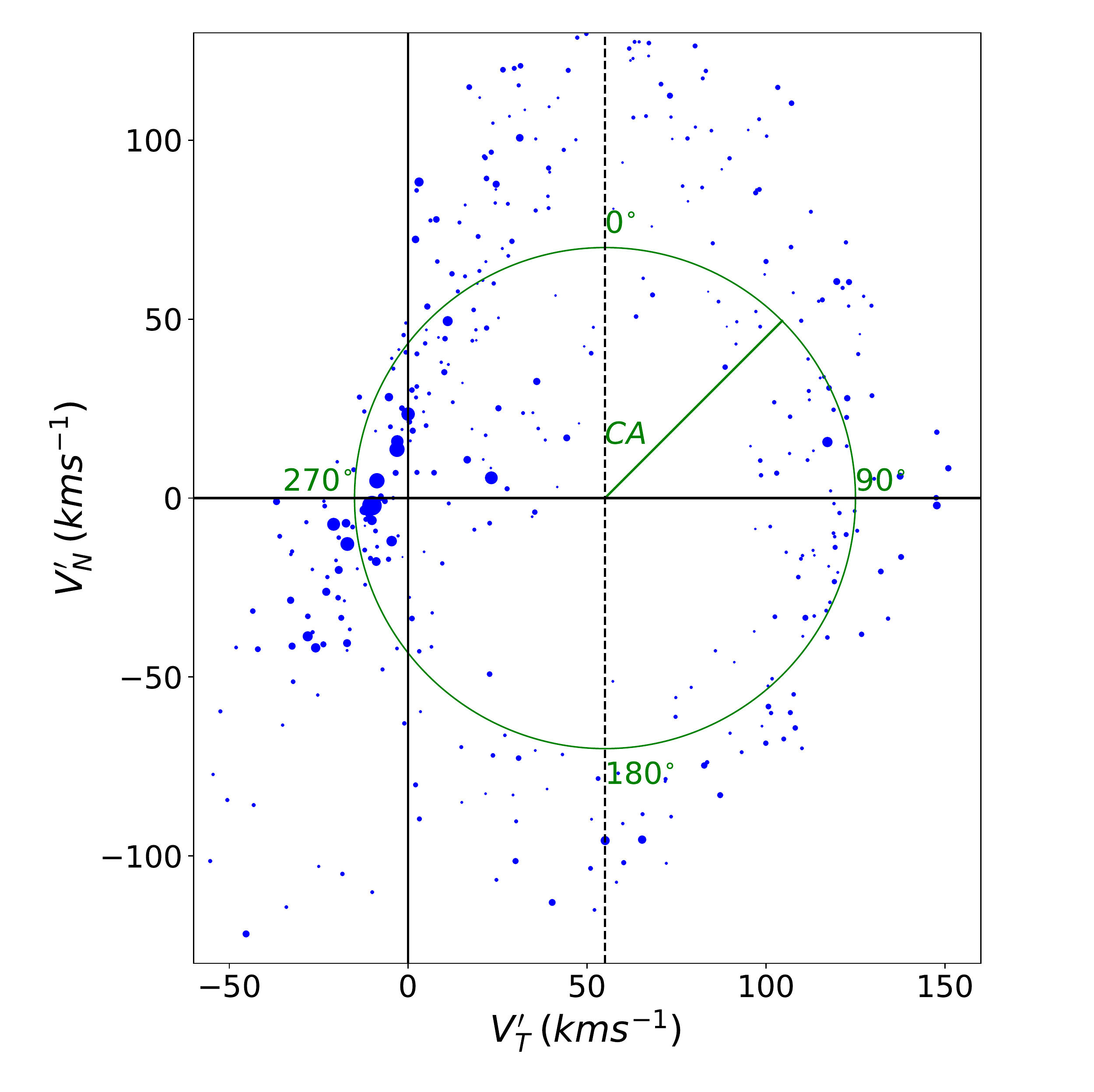}
\caption{T and N components of $\vec{V_{rel}^{\prime}}$ for all the switchbacks in the $\examplestream$ stream when rotated to coordinate system along $\alignsymbol = (-15\degr \alignT, 0\degr \alignN)$. They approximately lie on the a circle (green) when projected into the TN plane, limiting the direction through which the spacecraft can cut the structures. The area of the points is proportional to the switchback duration, highlighting that this is a function of clock angle (CA).}
          \label{fig:circle}%
\end{figure}

Ideally, the spacecraft would cut through the structure at many angles to build up a 3D shape. However, switchbacks are Alfv\'enic, meaning that magnetic and velocity perturbations are (anti-) correlated for (positive) negative polarity. Since Alfv\'enic fluctuations in the magnetic field lie on a sphere of constant radius, this corresponds to a sphere in velocity space, centred on some background velocity vector, with radius equal to the Alfv\'en speed \citep{Bruno2004, Matteini2014a, Matteini2015}. Switchbacks are deflections away from the background direction, and they rarely reach $180^{\degr}$ \citep{Woolley2020a}. Therefore, the possible values of $\left \langle \vec{V_{pl}}  \right \rangle$ create a torus in velocity space, which appears as a ring when projected in the TN plane, as demonstrated by Fig \ref{fig:circle}. This then severely limits the possible angles that the spacecraft can sample the structure. The position of this ring is determined by the background velocity and the spacecraft velocity with the size being determined by the Alfv\'en speed, which in turn is affected by the local magnetic field strength and density.

In order to determine the effect of changing $\alignsymbol$, we first consider a scenario where the ring of velocity is centred on $\alignsymbol$, as shown in Fig \ref{fig:cuts}(a). The ring of cutting angles can then be decomposed into the components $V_{rel, T}^{\prime}$ and $V_{rel, N}^{\prime}$, where $^{\prime}$ denotes a coordinate system rotated so $\alignsymbol$ is along R, as done in Fig \ref{fig:circle}. $V_{rel, T}^{\prime}$ and $V_{rel, N}^{\prime}$ follow a sine and cosine, respectively, as a function of ‘Clock Angle’ (CA). This is the clock face angle of the TN plane rotated to the Parker spiral, so $0^{\degr}$, $90^{\degr}$, $180^{\degr}$, $270^{\degr}$ correspond to +N, +T, -N, -T directions respectively. 

Using these components $\cutting$ is given by:
\begin{equation}\label{e:absangle}
\sin\cutting = \frac{\sqrt{V_{rel, T}^{\prime 2} + V_{rel, N}^{\prime 2}}}{|\vec{V_{rel}}|}.
\end{equation}

In this first example, $\cutting$ is constant for all clock angles, and therefore only one value of $D$ is measured. This can be seen explicitly as these components follow a sine and cosine, meaning that the numerator of Eq. \ref{e:absangle} is constant.

Fig \ref{fig:cuts}(b) shows the structure rotated by $\alignT$ in the RT plane. This introduces an offset in the $V_{rel,T}^{\prime}$ sine curve. This creates an asymmetry in $\cutting$ with CA when propagated through equation \ref{e:absangle}. This asymmetry is then reflected in $D$, which produces a range of durations seen in the time series data. In other words, the numerator of Eq. \ref{e:absangle} is now a function of CA.

The aim of this study is to find the alignment direction that was most consistent with long, thin structures, where the alignment direction is defined by the angle in the RT plane, $\alignT$, and TN plane, $\alignN$. This concept is more explicitly stated as the combination of $D$ and $\cutting$ that best followed Eq. \ref{e:duration}. The length of the structure along the spacecraft trajectory, $D$, and $\vec{V_{rel}}$ are observed quantities and do not depend on $\alignsymbol$. However, this is not true for the cutting angle, $\cutting$, which is the angle between $\vec{V_{rel}}$ and $\alignsymbol$. Therefore, each alignment direction we tested created a new set of $\cutting$, while $D$ remained the same. 

We identified switchbacks as structures where the magnetic field deflected more than $45\degr$ away from the Parker spiral direction. We recognise that the start and end of the switchback is not when the structure deflects above this threshold. Therefore, we measured the duration of the identified switchback structure from when the deflection was above $30\degr$, to best capture the full duration of the structures. The Parker spiral direction was calculated using twelve hour modes of the solar wind parameters, in an attempt to remove the switchback contribution \citep{Bale2019}. Successive switchbacks were merged if they were separated by less than 2 seconds.

In order to determine whether with the combination of $\cutting$ and $D$ followed Eq. \ref{e:duration}, we first binned the switchback data for a given stream into $2\degr$ bins in $\cutting$. A minimum number of 10 points per bin was enforced in an attempt to remove statistically insignificant bins. Eq. \ref{e:duration} only applies to cuts made through the centre of the structure. Therefore, we reduced the dataset to those switchbacks that are most likely to be cuts through the centre, by finding the maximum value of $D$ for each bin. This removed situations where the spacecraft clipped the edge of the structure or encountered substructure that was not relevant to the average switchback shape.

This procedure allowed Eq. \ref{e:duration} to be fitted to $D$ as a function of $\cutting$ for a stream, where the free parameters are $W$ and $\hypot$. A least squares algorithm was used, with constraints that $W > 0 \, \rm{km}$ and $0\degr < \hypot < 90\degr$ and a minimum of five bins in $\cutting$ were used. Functionally, $\hypot$ means that the longest switchbacks do not have to lie at $0\degr$ $\cutting$ and also provided an estimate of aspect ratio. The $R^{2}$ value of the fit was used to quantify the goodness of fit of the data to Eq. \ref{e:duration}. This in turn allows for a determination of how consistent a particular alignment direction is with the underlying switchback direction for that stream. For each stream, the values of $W$ and $\hypot$ were selected so that the value of $R^{2}$ was maximised. Of the fifteen streams outlined in Fig \ref{fig:streamsummary}, six streams produced robust results that are discussed in the next section.

\section{Results and discussion}\label{sec:results}

\begin{figure}[ht!]
\centering
\includegraphics[width=\hsize]{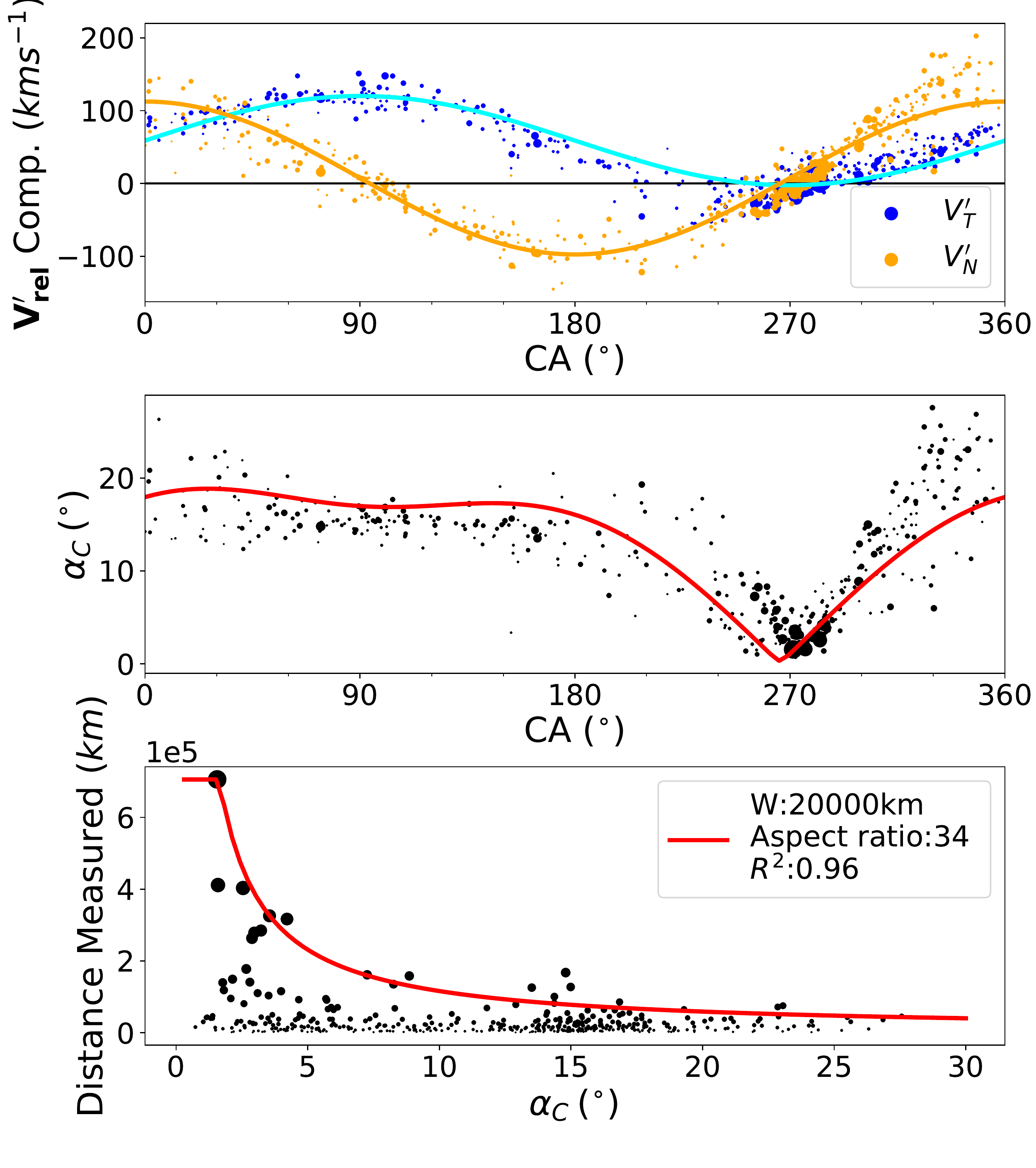}
\caption{Switchbacks for the $\examplestream$ stream are shown to be consistent with long, thin structures along $\alignsymbol = (-15\degr \alignT, 0\degr \alignN)$. The fitting procedure in Section \ref{sec:method} gave values of $W = 20,000 \, \rm{km}$ and a minimum aspect ratio of $34$. The area of the scatter points are proportional to the distance measured at by the spacecraft.}
          \label{fig:example}%
\end{figure}

The results for the $\examplestream$ stream are shown in Fig \ref{fig:example}, with a best fit alignment direction of $(\alignT = -15\degr, \alignN = 0\degr)$ that was found using the method outlined in Sec \ref{sec:method}. The top panel shows the $V_{rel, T}^{\prime}$ and $V_{rel, N}^{\prime}$ components, which follow a sine and cosine function as a function of CA that are offset from $0 \, kms^{-1}$. This is simply unwrapping the ring presented in Fig \ref{fig:circle}, with respect to the origin ($V_{rel, T}^{\prime}$=$0 \, kms^{-1}$ , $V_{rel, N}^{\prime}$ = $0 \, kms^{-1}$ ). The relationship between $\cutting$ and CA (middle panel) was calculated using Eq. \ref{e:absangle}, which produces a correlation between lower $\cutting$ and longer $D$. This is more explicitly demonstrated in the bottom panel where $D$ is plotted as a function of $\cutting$, with a best fit (red line) having a width of $20,000 \, \rm{km}$ and an aspect ratio of 34. 

\begin{figure*}[!t]
\centering
\includegraphics[width=\textwidth]{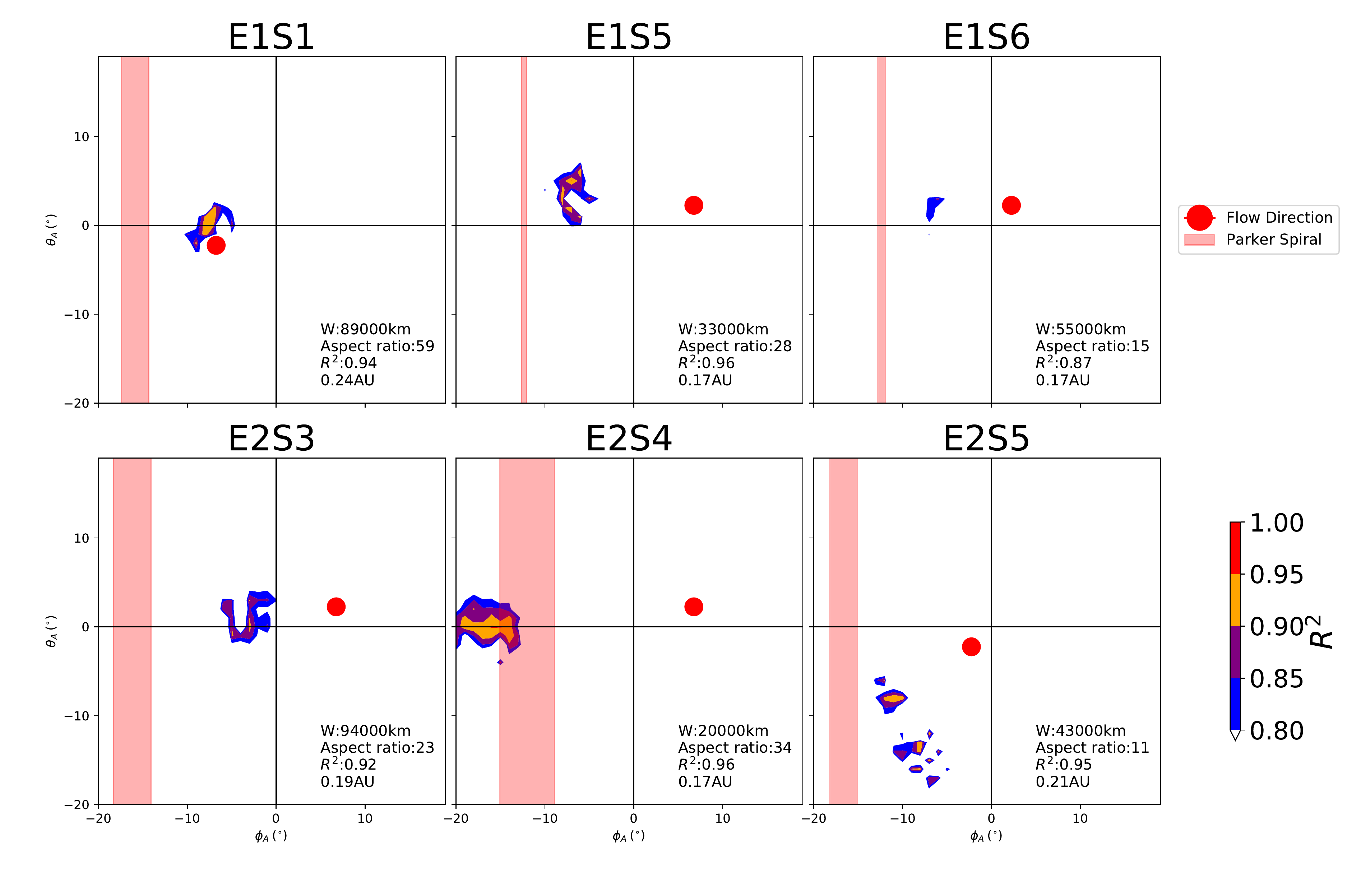}
\caption{Results of varying alignment direction, defined by the angle in the RT plane, $\alignT$, and TN plane, $\alignN$. The radial direction is the centre of the black cross hairs, with the Parker spiral and background flow direction also shown. The background flow direction is taken as the deHoffman-Teller frame (see \citet{Horbury2020a}), rather than the average velocity per stream which is influenced by the switchbacks. Higher scores correspond to an alignment direction most consistent with long, thin structures. This is always offset in -T direction, roughly consistent with Parker spiral direction calculated using a twelve hour mode of velocity.}
          \label{fig:global}%
\end{figure*}

The width is a robustly determined quantity, with $W$ in different streams varying from $20,000 \, \rm{km}$ to $94,000 \, \rm{km}$ with a mean of $56,000 \, \rm{km}$. We note that $W$ appears to increase as the spacecraft moves further from the Sun (Fig \ref{fig:global}), although there is not a large enough spread in distance from the Sun to determine the rate of expansion. The aspect ratio ranges from $11$ to $59$ with a mean value of $28$, although this parameter is only a lower bound, so should only be considered as being of the order of 10. Using these two average parameters the length of the structures, $L$, is $\sim 500,000~ \, \rm{km}$ which is of the order of a solar radius.

The result of fitting Eq. \ref{e:duration} to stream $\examplestream$ produced an $R^{2}$ of $0.96$. The dependence of $R^{2}$ (only values greater than 0.8 are shown) with alignment direction can be seen in Fig \ref{fig:global} for several solar wind streams. Each stream displays a tight grouping in $R^{2}$, indicating that there was a unique alignment direction. This grouping always has a negative $\alignT$, close to the local Parker spiral direction (red shading) which was also typically larger in amplitude than $\alignN$. In contrast, there was no correlation between switchback alignment direction and the background solar wind flow direction (red dot). This suggests that the average switchback alignment direction is not related to the +T solar wind flow reported by \citet{Kasper2019}. This statement does not rule out switchbacks as significant contributors to solar wind flow deflections, only that their longest axis is not aligned along the flow deflection. 

However, this analysis does raise a subtlety when trying to interpret how switchbacks may be related to the background flow of the solar wind. The direction of magnetic deflection (which we discuss as clock angle) within each switchback affects the way in which the spacecraft cuts through it, since magnetic and velocity fluctuations are correlated. Therefore, the direction of magnetic deflection determines the apparent duration seen by the spacecraft, meaning that duration alone can not be used as a proxy for the true physical size of a switchback. In other words, it would be false to argue that if an individual switchback has a longer duration at a spacecraft, it must be physically larger and therefore make a more significant contribution to the solar wind than one which is shorter. A more accurate approach would be to consider some metric that does not depend on duration, such as magnetic curvature or the Poynting flux \citep{Mozer2020, Woolley2020a}. Both authors showed that the radial Poynting flux depends on the angle of magnetic deflection rather than duration, with a maximum reached at $90\degr$ from radial.

This relationship between deflection direction and apparent duration also means that the spacecraft data will include a sampling bias in the direction of the longest switchbacks' deflection. A simple average of all data within a stream will include a greater contribution from longer switchbacks. A consequence of this is that the average velocity direction will point along the deflection direction of the longest switchbacks, meaning that care must be taken when considering the average properties of a solar wind stream.

The analysis in Sec \ref{sec:method} does not work for every stream, where either a pattern in duration could not be identified or there were not enough points to draw a solid conclusion. We note that this method may only apply in PSP's first three encounters, as these are times when the spacecraft was approximately co-rotating with the Sun's surface. As a result, plasma from the same source region was sampled for several hours \citep{Bale2019}, which allowed a large number of switchbacks to be observed per stream. Although PSP will pass through two co-rotation points per orbit, this window of co-rotation will become smaller with each orbit, reducing the number of switchbacks measured from the same source region. This would mean that the analysis in this paper may not be possible for PSP in future orbits, or indeed for other spacecraft at 1AU.

We can nevertheless make an estimate of how switchbacks should appear in different situations in the heliosphere. Here we assume a constant aspect ratio, since we could not accurately determine how this varied with radial distance. For PSP's future orbits, if switchbacks are Parker spiral aligned then they will become more radial as distance to the Sun decreases, while the width would decrease. This, coupled with the greater spacecraft transverse velocity, means that the spacecraft will sample the structure at greater angles to their longest direction. Combined with their high aspect ratio, this means that the distribution of switchback duration will be more skewed towards shorter durations.

For larger radial distances from the Sun, the switchback alignment will be further from the radial direction. This, along with a lower spacecraft transverse velocity, will again lead to the spacecraft travelling through the structure at larger angles. This should lead to longer duration switchbacks becoming more rare, although the rate at which switchback width increases could offset this. This was the case for Helios measurements where most velocity enhancements appeared as a single data point in measurements with a cadence of 40.5 seconds \citep{Horbury2018a}. These are broad predictions as the true answer relies on a complex interplay between: spacecraft speed; Alfv\'en speed; Parker spiral direction; background flow deflection; and the evolution of the switchbacks with distance.

\section{Conclusion}
We have investigated how the direction at which the spacecraft cuts through switchbacks influences the apparent duration that is seen in the time series data. We assumed that switchbacks are long, thin structures with cylindrical symmetry that are aligned in a certain direction, as suggested by \citet{Horbury2020a}. We then searched for the average alignment direction of switchbacks within several solar wind streams at $\sim 35-50~R_{S}$, and compared this to known physical directions. We found that this alignment direction was always away from radial, by $\sim 10\degr$, with the same sense as the Parker spiral field. Although our assumptions cannot be proven with single spacecraft measurements, they produced consistent results repeated over six solar wind streams identified across the first two PSP encounters. The average width of the switchbacks, assuming cylindrical symmetry, was of the order of $50,000 \, \rm{km}$, with this width increasing as the spacecraft moved further from Sun. The aspect ratio of these structures was of order 10. 

Our results show that each solar wind stream has its own average alignment direction, although this was similar across streams. Since PSP was co-rotating near perihelion, this may imply that differences in alignment direction between streams relates to the magnetic orientation of the source region -- but this could also be due to switchbacks being instead guided by the local interplanetary magnetic field. If switchbacks are related to coronal jets, like those simulated by \citet{Roberts2018}, it is more likely that an Alfv\'en wave packet launched by the jet would survive out to PSP, rather than the jet itself as discussed in \citet{Sterling2020}. The authors suggest that the length of such a wave-packet would be $\sim 600,000 \, \rm{km}$, similar to that from our results. It may be that a patch of switchbacks are the result of the spacecraft travelling inside a flux tube with coronal transients released intermittently at the base, with the quiet radial solar wind between these patches being the ambient solar wind predicted by \citet{Parker1958}. Our results do not rule out other possibilities for switchback origin however, such as the evolution of Alfv\'enic fluctuations \citep{Squire2020}.

We also remark that dips in the field magnitude associated with sharp magnetic deflections at the edge of switchbacks \citep{Krasnoselskikh2020a, Agapitov2020}, along with their arc-polarised behaviour, are reminiscent of phase steepened Alfv\'en waves studied by Tsurutani and others \citep{Tsurutani1994a, Vasquez1996a, Tsurutani2002, Tsurutani2005}. While it is not true of all switchbacks, there was a subset that demonstrated a slow rotation in the magnetic field followed by a sharp change back to the original field direction, and vice versa. A preliminary survey showed consistency of deflection direction for those switchbacks that exhibit sharp boundaries. This may suggest that these cases are linked to how the spacecraft cuts through the structure,with the sharp discontinuities being a part of the switchback structure, rather than a sampling effect. We intend to revisit this subject in a future study.

This analysis raises the point that the switchback duration in time series data is partly a consequence of spacecraft motion with respect to the switchback structures, rather than just a reflection of their true physical size. This means that it would be wrong to argue that larger duration switchbacks make a more significant contribution to the solar wind than those that are shorter. Indeed, it may be that those considered here, from PSP's first two perihelia while the spacecraft was near co-rotation, have longer durations at the spacecraft than those measured during later orbits when the spacecraft is moving considerably faster.

\begin{acknowledgements}
RL was supported by an Imperial College President's Scholarship, TSH by STFC ST/S000364/1, TW by ST/N504336/1, LDW by ST/S000364/1. SDB acknowledges the support of the Leverhulme Trust Visiting Professor program. S.T.B. was supported by NASA Headquarters under the NASA Earth and Space Science Fellowship Program Grant 80NSSC18K1201. The SWEAP and FIELDS teams acknowledge support from NASA contract NNN06AA01C. This work has made use of the open source and free community-developed space physics packages HelioPy \citep{heliopy2020} and SpiceyPy \citep{spiceypy2020}.
\end{acknowledgements}

% WARNING
%-------------------------------------------------------------------
% Please note that we have included the references to the file aa.dem in
% order to compile it, but we ask you to:
%
% - use BibTeX with the regular commands:
%   \bibliographystyle{aa} % style aa.bst
%   \bibliography{Yourfile} % your references Yourfile.bib
%
% - join the .bib files when you upload your source files
%-------------------------------------------------------------------
\bibliographystyle{aa} % style aa.bst
\bibliography{library.bib} % your references Yourfile.bib

\begin{thebibliography}{39}
\expandafter\ifx\csname natexlab\endcsname\relax\def\natexlab#1{#1}\fi

\bibitem[{Agapitov {et~al.}(2020)Agapitov, {Dudok de Wit}, Mozer, Bonnell,
  Drake, Malaspina, Krasnoselskikh, Bale, Whittlesey1, Case, Chaston, Froment,
  Goetz10, Goodrich, Harvey, Kasper, Korreck, Larson, Livi, MacDowall, Pulupa,
  Revillet, Stevens, \& Wygant}]{Agapitov2020}
Agapitov, O.~V., {Dudok de Wit}, T., Mozer, F.~S., {et~al.} 2020, The
  Astrophysical Journal, 891

\bibitem[{Annex {et~al.}(2020)Annex, Pearson, Seignovert, Carcich, Eichhorn,
  Mapel, von Forstner, McAuliffe, del Rio, Berry, Aye, Stefko, de~Val-Borro,
  Kulumani, Murakami, Niemeyer, \& Medley}]{spiceypy2020}
Annex, A., Pearson, B., Seignovert, B., {et~al.} 2020, {AndrewAnnex/SpiceyPy:
  SpiceyPy 3.1.1}

\bibitem[{Badman {et~al.}(2020)Badman, Bale, Mart, Pansenco, Velli, Bonnell,
  Case, Stansby, Buitrago-casas, Victor, Wit, Goetz, Harvey, Kasper, Korreck,
  Stevens, \& Whittlesey}]{Badman2020}
Badman, S.~T., Bale, S.~D., Mart, J.~C., {et~al.} 2020, Astrophysical Journal,
  246

\bibitem[{Bale {et~al.}(2019)Bale, Badman, Bonnell, Bowen, Burgess, Case,
  Cattell, Chandran, Chaston, Chen, Drake, de~Wit, Eastwood, Ergun, Farrell,
  Fong, Goetz, Goldstein, Goodrich, Harvey, Horbury, Howes, Kasper, Kellogg,
  Klimchuk, Korreck, Krasnoselskikh, Krucker, Laker, Larson, MacDowall,
  Maksimovic, Malaspina, Martinez-Oliveros, McComas, Meyer-Vernet, Moncuquet,
  Mozer, Phan, Pulupa, Raouafi, Salem, Stansby, Stevens, Szabo, Velli, Woolley,
  \& Wygant}]{Bale2019}
Bale, S.~D., Badman, S.~T., Bonnell, J.~W., {et~al.} 2019, Nature, 576, 237

\bibitem[{Bale {et~al.}(2016)Bale, Goetz, Harvey, Turin, Bonnell, {Dudok de
  Wit}, Ergun, MacDowall, Pulupa, Andre, Bolton, Bougeret, Bowen, Burgess,
  Cattell, Chandran, Chaston, Chen, Choi, Connerney, Cranmer, Diaz-Aguado,
  Donakowski, Drake, Farrell, Fergeau, Fermin, Fischer, Fox, Glaser, Goldstein,
  Gordon, Hanson, Harris, Hayes, Hinze, Hollweg, Horbury, Howard, Hoxie,
  Jannet, Karlsson, Kasper, Kellogg, Kien, Klimchuk, Krasnoselskikh, Krucker,
  Lynch, Maksimovic, Malaspina, Marker, Martin, Martinez-Oliveros, McCauley,
  McComas, McDonald, Meyer-Vernet, Moncuquet, Monson, Mozer, Murphy, Odom,
  Oliverson, Olson, Parker, Pankow, Phan, Quataert, Quinn, Ruplin, Salem,
  Seitz, Sheppard, Siy, Stevens, Summers, Szabo, Timofeeva, Vaivads, Velli,
  Yehle, Werthimer, \& Wygant}]{Bale2016}
Bale, S.~D., Goetz, K., Harvey, P.~R., {et~al.} 2016, Space Science Reviews,
  204, 49

\bibitem[{Balogh {et~al.}(1999)Balogh, Forsyth, Lucek, Horbury, \&
  Smith}]{Balogh1999}
Balogh, A., Forsyth, R.~J., Lucek, E.~A., Horbury, T.~S., \& Smith, E.~J. 1999,
  Geophysical Research Letters, 26, 631

\bibitem[{Bowen {et~al.}(2020)Bowen, Bale, Bonnell, Larson, Mallet, McManus,
  Mozer, Pulupa, Vasko, \& Verniero}]{Bowen2020}
Bowen, T.~A., Bale, S.~D., Bonnell, J.~W., {et~al.} 2020, The Astrophysical
  Journal, 899, 74

\bibitem[{Bruno {et~al.}(2004)Bruno, Carbone, Primavera, Malara, Sorriso-Valvo,
  Bavassano, \& Veltri}]{Bruno2004}
Bruno, R., Carbone, V., Primavera, L., {et~al.} 2004, Annales Geophysicae, 22,
  3751

\bibitem[{Case {et~al.}(2020)Case, Kasper, Stevens, Korreck, Paulson, Daigneau,
  Caldwell, Freeman, Henry, Klingensmith, Bookbinder, Robinson, Berg, Tiu,
  Wright, Reinhart, Curtis, Ludlam, Larson, Whittlesey, Livi, Klein, \&
  Martinovi{\'{c}}}]{Case2020}
Case, A.~W., Kasper, J.~C., Stevens, M.~L., {et~al.} 2020, The Astrophysical
  Journal Supplement Series, 246, 43

\bibitem[{{De Koning} {et~al.}(2009){De Koning}, Pizzo, \&
  Biesecker}]{DeKoning2009}
{De Koning}, C.~A., Pizzo, V.~J., \& Biesecker, D.~A. 2009, Solar Phys, 256,
  167

\bibitem[{{Dudok de Wit} {et~al.}(2020){Dudok de Wit}, Krasnoselskikh, Bale,
  Bonnell, Bowen, Chen, Froment, Goetz, Harvey, Jagarlamudi, Larosa, MacDowall,
  Malaspina, Matthaeus, Pulupa, Velli, \& Whittlesey}]{DudokdeWit2020}
{Dudok de Wit}, T., Krasnoselskikh, V.~V., Bale, S.~D., {et~al.} 2020, The
  Astrophysical Journal Supplement Series, 246, 39

\bibitem[{Horbury {et~al.}(2018)Horbury, Matteini, \& Stansby}]{Horbury2018a}
Horbury, T.~S., Matteini, L., \& Stansby, D. 2018, Monthly Notices of the Royal
  Astronomical Society, 478, 1980

\bibitem[{Horbury {et~al.}(2020)Horbury, Woolley, Laker, Matteini, Eastwood,
  Bale, Velli, Chandran, Phan, Raouafi, Goetz, Harvey, Pulupa, Klein, {Dudok de
  Wit}, Kasper, Korreck, Case, Stevens, Whittlesey, Larson, MacDowall,
  Malaspina, \& Livi}]{Horbury2020a}
Horbury, T.~S., Woolley, T., Laker, R., {et~al.} 2020, The Astrophysical
  Journal Supplement Series, 246, 45

\bibitem[{Kasper {et~al.}(2016)Kasper, Abiad, Austin, Balat-Pichelin, Bale,
  Belcher, Berg, Bergner, Berthomier, Bookbinder, Brodu, Caldwell, Case,
  Chandran, Cheimets, Cirtain, Cranmer, Curtis, Daigneau, Dalton, Dasgupta,
  DeTomaso, Diaz-Aguado, Djordjevic, Donaskowski, Effinger, Florinski, Fox,
  Freeman, Gallagher, Gary, Gauron, Gates, Goldstein, Golub, Gordon, Gurnee,
  Guth, Halekas, Hatch, Heerikuisen, Ho, Hu, Johnson, Jordan, Korreck, Larson,
  Lazarus, Li, Livi, Ludlam, Maksimovic, McFadden, Marchant, Maruca, McComas,
  Messina, Mercer, Park, Peddie, Pogorelov, Reinhart, Richardson, Robinson,
  Rosen, Skoug, Slagle, Steinberg, Stevens, Szabo, Taylor, Tiu, Turin, Velli,
  Webb, Whittlesey, Wright, Wu, \& Zank}]{Kasper2016}
Kasper, J.~C., Abiad, R., Austin, G., {et~al.} 2016, Space Science Reviews,
  204, 131

\bibitem[{Kasper {et~al.}(2019)Kasper, Bale, Belcher, Berthomier, Case,
  Chandran, Curtis, Gallagher, Gary, Golub, Halekas, Ho, Horbury, Hu, Huang,
  Klein, Korreck, Larson, Livi, Maruca, Lavraud, Louarn, Maksimovic,
  Martinovic, McGinnis, Pogorelov, Richardson, Skoug, Steinberg, Stevens,
  Szabo, Velli, Whittlesey, Wright, Zank, MacDowall, McComas, McNutt, Pulupa,
  Raouafi, \& Schwadron}]{Kasper2019}
Kasper, J.~C., Bale, S.~D., Belcher, J.~W., {et~al.} 2019, Nature, 576, 228

\bibitem[{Krasnoselskikh {et~al.}(2020)Krasnoselskikh, Larosa, Agapitov,
  de~Wit, Moncuquet, Mozer, Stevens, Bale, Bonnell, Froment, Goetz, Goodrich,
  Harvey, Kasper, MacDowall, Malaspina, Pulupa, Raouafi, Revillet, Velli, \&
  Wygant}]{Krasnoselskikh2020a}
Krasnoselskikh, V., Larosa, A., Agapitov, O., {et~al.} 2020, The Astrophysical
  Journal, 893, 93

\bibitem[{Liewer {et~al.}(2009)Liewer, {De Jong}, Hall, Howard, Thompson,
  Culhane, Bone, \& {Van Driel-Gesztelyi}}]{Liewer2009}
Liewer, P.~C., {De Jong}, E.~M., Hall, J.~R., {et~al.} 2009, Solar Physics,
  256, 57

\bibitem[{Matteini {et~al.}(2014)Matteini, Horbury, Neugebauer, \&
  Goldstein}]{Matteini2014a}
Matteini, L., Horbury, T.~S., Neugebauer, M., \& Goldstein, B.~E. 2014,
  Geophysical Research Letters, 41, 259

\bibitem[{Matteini {et~al.}(2015)Matteini, Horbury, Pantellini, Velli, \&
  Schwartz}]{Matteini2015}
Matteini, L., Horbury, T.~S., Pantellini, F., Velli, M., \& Schwartz, S.~J.
  2015, Astrophysical Journal, 802, 11

\bibitem[{Mcmanus {et~al.}(2020)Mcmanus, Bowen, Mallet, Chen, Chandran, Bale,
  Larson, {Dudok De Wit}, Kasper, Stevens, Whittlesey, Livi, Korreck, Goetz,
  Harvey, Pulupa, Macdowall, Malaspina, Case, \& Bonnell}]{Mcmanus2020}
Mcmanus, M.~D., Bowen, T.~A., Mallet, A., {et~al.} 2020, The Astrophysical
  Journal Supplement Series, 246, 67

\bibitem[{Mozer {et~al.}(2020)Mozer, Agapitov, Bale, Bonnell, Case, Chaston,
  Curtis, de~Wit, Goetz, Goodrich, Harvey, Kasper, Korreck, Krasnoselskikh,
  Larson, Livi, MacDowall, Malaspina, Pulupa, Stevens, Whittlesey, \&
  Wygant}]{Mozer2020}
Mozer, F.~S., Agapitov, O.~V., Bale, S.~D., {et~al.} 2020, The Astrophysical
  Journal Supplement Series, 246, 68

\bibitem[{Neugebauer \& Goldstein(2013)}]{Neugebauer2013}
Neugebauer, M. \& Goldstein, B.~E. 2013, in AIP Conference Proceedings, Vol.
  1539, 46--49

\bibitem[{Parker(1958)}]{Parker1958}
Parker, E.~N. 1958, The Astrophysical Journal, 128, 664

\bibitem[{Raouafi {et~al.}(2016)Raouafi, Patsourakos, Pariat, Young, Sterling,
  \& Raouafi}]{Raouafi2016}
Raouafi, N.~E., Patsourakos, {\textperiodcentered}.~S., Pariat,
  {\textperiodcentered}.~E., {et~al.} 2016, Space Sci Rev, 201, 1

\bibitem[{Roberts {et~al.}(2018)Roberts, Uritsky, DeVore, \&
  Karpen}]{Roberts2018}
Roberts, M.~A., Uritsky, V.~M., DeVore, C.~R., \& Karpen, J.~T. 2018, The
  Astrophysical Journal, 866, 14

\bibitem[{Rouillard {et~al.}(2020)Rouillard, Kouloumvakos, Vourlidas, Kasper,
  Bale, Raouafi, Lavraud, Howard, Stenborg, Stevens, Poirier, Davies, Hess,
  Higginson, Lavarra, Viall, Korreck, Pinto, Griton, R{\'{e}}ville, Louarn, Wu,
  Dalmasse, G{\'{e}}not, Case, Whittlesey, Larson, Halekas, Livi, Goetz,
  Harvey, Macdowall, Malaspina, Pulupa, Bonnell, {Dudok De Witt}, \&
  Penou}]{Rouillard2020a}
Rouillard, A.~P., Kouloumvakos, A., Vourlidas, A., {et~al.} 2020, The
  Astrophysical Journal Supplement Series, 246

\bibitem[{Squire {et~al.}(2020{\natexlab{a}})Squire, Chandran, \&
  Meyrand}]{Squire2020a}
Squire, J., Chandran, B. D.~G., \& Meyrand, R. 2020{\natexlab{a}}, The
  Astrophysical Journal Letters, 891, L2

\bibitem[{Squire {et~al.}(2020{\natexlab{b}})Squire, Chandran, \&
  Meyrand}]{Squire2020}
Squire, J., Chandran, B. D.~G., \& Meyrand, R. 2020{\natexlab{b}}, 2, 1

\bibitem[{Stansby {et~al.}(2020)Stansby, Rai, Argall, JeffreyBroll,
  Haythornthwaite, Erwin, Shaw, Aditya, Saha, Ireland, Lim, Badman, Mishra,
  Badger, \& DupuisIRT}]{heliopy2020}
Stansby, D., Rai, Y., Argall, M., {et~al.} 2020, {heliopython/heliopy: HelioPy
  0.13.0}

\bibitem[{Sterling \& Moore(2020)}]{Sterling2020}
Sterling, A.~C. \& Moore, R.~L. 2020, The Astrophysical Journal Letters, 896

\bibitem[{Szabo {et~al.}(2020)Szabo, Larson, Whittlesey, Stevens, Lavraud,
  Phan, Wallace, Jones-Mecholsky, Arge, Badman, Odstrcil, Pogorelov, Kim,
  Riley, Henney, Bale, Bonnell, Case, {Dudok De Wit}, Goetz, Harvey, Kasper,
  Korreck, Koval, Livi, Macdowall, Malaspina, \& Pulupa}]{Szabo2020}
Szabo, A., Larson, D., Whittlesey, P., {et~al.} 2020, The Astrophysical Journal
  Supplement Series, 246, 47

\bibitem[{Tenerani {et~al.}(2020)Tenerani, Velli, Matteini, R{\'{e}}ville, Shi,
  Bale, Kasper, Bonnell, Case, {Dudok De Wit}, Goetz, Harvey, Klein, Korreck,
  Larson, Livi, Macdowall, Malaspina, Pulupa, Stevens, \&
  Whittlesey}]{Tenerani2020}
Tenerani, A., Velli, M., Matteini, L., {et~al.} 2020, The Astrophysical Journal
  Supplement Series, 246, 32

\bibitem[{Tsurutani {et~al.}(2002)Tsurutani, Dasgupta, Galvan, Neugebauer,
  Lakhina, Arballo, Winterhalter, Goldstein, \& Buti}]{Tsurutani2002}
Tsurutani, B.~T., Dasgupta, B., Galvan, C., {et~al.} 2002, Geophysical Research
  Letters, 29, 86

\bibitem[{Tsurutani {et~al.}(1994)Tsurutani, Ho, Smith, Neugebauer, Goldstein,
  Mok, Arballo, Balogh, Southwood, \& Feldman}]{Tsurutani1994a}
Tsurutani, B.~T., Ho, C.~M., Smith, E.~J., {et~al.} 1994, Geophysical Research
  Letters, 21, 2267

\bibitem[{Tsurutani {et~al.}(2005)Tsurutani, Lakhina, Pickett, Guarnieri, Lin,
  \& Goldstein}]{Tsurutani2005}
Tsurutani, B.~T., Lakhina, G.~S., Pickett, J.~S., {et~al.} 2005, Nonlinear
  Processes in Geophysics, 12, 321

\bibitem[{Vasquez \& Hollweg(1996)}]{Vasquez1996a}
Vasquez, B.~J. \& Hollweg, J.~V. 1996, Journal of Geophysical Research: Space
  Physics, 101, 13527

\bibitem[{Verniero {et~al.}(2020)Verniero, Larson, Livi, Rahmati, McManus,
  Pyakurel, Klein, Bowen, Bonnell, Alterman, Whittlesey, Malaspina, Bale,
  Kasper, Case, Goetz, Harvey, Korreck, MacDowall, Pulupa, Stevens, \&
  de~Wit}]{Verniero2020}
Verniero, J.~L., Larson, D.~E., Livi, R., {et~al.} 2020, The Astrophysical
  Journal Supplement Series, 248, 5

\bibitem[{Woodham {et~al.}(2020)Woodham, Horbury, Matteini, Woolley, Laker,
  Bale, Nicolaou, Stawarz, Stansby, Hietala, Larson, Livi, Verniero, McManus,
  Kasper, Korreck, \& Raouafi}]{Woodham2020}
Woodham, L., Horbury, T.~S., Matteini, L., {et~al.} 2020, Astronomy {\&}
  Astrophysics

\bibitem[{Woolley {et~al.}(2020)Woolley, Matteini, Horbury, Bale, Woodham,
  Laker, Alterman, Bonnell, Case, Kasper, Klein, Martinovi{\'{c}}, \&
  Stevens}]{Woolley2020a}
Woolley, T., Matteini, L., Horbury, T.~S., {et~al.} 2020, Monthly Notices of
  the Royal Astronomical Society, 498, 5524

\end{thebibliography}

\onecolumn

\begin{appendix}
\section{Stream Summary}

% Please add the following required packages to your document preamble:
% \usepackage{graphicx}
\begin{table}[h]

\caption{Summarising information for the streams used in this study including dates and number of switchbacks, with the results only shown for those streams where the analysis in Sec \ref{sec:method} was successful. A graphical representation of these streams can be seen in Fig \ref{fig:streamsummary}. The Radial Distance ($R$) is the mean distance from the Sun within each stream. The source region is shown, with CH standing for coronal hole and Inside and Outside refer to the streamer belt.}\label{tab:1}

\begin{tabular}{ccccccccccc}
\hline\hline    
ID    & Start            & End              & Number & Source  & R $(AU)$ & $\phi_{A} \, (\degr)$ & $\theta_{A}  \, (\degr)$ & $R^{2}$ & $W \, (km)$ & Aspect Ratio \\ \hline
E1S1  & 31/10/2018 01:20 & 01/11/2018 10:53 & 239             & CH      & 0.24                   & -8                    & 0                        & 0.94    & 89000       & 59           \\
E1S2  & 01/11/2018 16:56 & 02/11/2018 11:28 & 114             & CH      & 0.22                   & -                     & -                        & -       & -           & -            \\
E1S3  & 02/11/2018 11:35 & 03/11/2018 19:37 & 245             & CH      & 0.2                    & -                     & -                        & -       & -           & -            \\
E1S4  & 03/11/2018 20:13 & 04/11/2018 18:35 & 283             & CH      & 0.18                   & -                     & -                        & -       & -           & -            \\
E1S5  & 05/11/2018 00:48 & 05/11/2018 20:30 & 223             & CH      & 0.17                   & -7                    & 2                        & 0.96    & 33000       & 28           \\
E1S6  & 05/11/2018 22:28 & 07/11/2018 09:30 & 510             & CH      & 0.17                   & -6                    & 3                        & 0.87    & 55000       & 15           \\
E1S7  & 07/11/2018 09:54 & 08/11/2018 08:00 & 375             & CH      & 0.18                   & -                     & -                        & -       & -           & -            \\
E1S8  & 08/11/2018 17:50 & 09/11/2018 06:19 & 178             & CH      & 0.19                   & -                     & -                        & -       & -           & -            \\
E1S9  & 09/11/2018 14:23 & 10/11/2018 12:30 & 330             & CH      & 0.21                   & -                     & -                        & -       & -           & -            \\
E1S10 & 15/11/2018 17:23 & 19/11/2018 10:39 & 1054            & CH      & 0.38                   & -                     & -                        & -       & -           & -            \\
E2S1  & 30/03/2019 00:52 & 30/03/2019 18:36 & 214             & NA      & 0.25                   & -                     & -                        & -       & -           & -            \\
E2S2  & 30/03/2019 21:24 & 01/04/2019 08:30 & 295             & NA      & 0.22                   & -                     & -                        & -       & -           & -            \\
E2S3  & 01/04/2019 09:30 & 03/04/2019 08:00 & 285             & Inside  & 0.19                   & -3                    & 0                        & 0.92    & 94000       & 23           \\
E2S4  & 03/04/2019 08:52 & 06/04/2019 11:49 & 435             & Outside & 0.17                   & -15                   & 0                        & 0.96    & 20000       & 34           \\
E2S5  & 06/04/2019 22:00 & 07/04/2019 18:19 & 198             & Outside & 0.19                   & -                     & -                        & -       & -           & -            \\
E2S6  & 08/04/2019 03:50 & 09/04/2019 19:09 & 396             & Inside  & 0.21                   & -7                    & -15                      & 0.95    & 43000       & 11           \\ \hline
\end{tabular}%

\end{table}

%\caption[\textwidth]{Summarising information for the streams used in this study, with the results only shown for those streams where the analysis in Sec \ref{sec:method} was successful. The Radial Distance is the mean distance from the Sun.}
%\label{tab:my-table}
\end{appendix}

\end{document}